\documentclass[12pt]{article}
\usepackage{epsfig}

\usepackage{a4}

\usepackage{amsmath}

\begin{document}
\title{A remark on black holes of Chern-Simons gravities in $2n+1$ dimensions: $n=1,2,3$}
\author{{\large }
{\large D. H. Tchrakian}$^{\ddagger \star}$ \\ \\
$^{\ddagger}${\small School of Theoretical Physics, Dublin Institute for Advanced Studies,
10 Burlington Road, Dublin 4, Ireland}\\
$^{\star}${\small Department of Computer Science, Maynooth University, Maynooth, Ireland}}

\date{}
\newcommand{\dd}{\mbox{d}}
\newcommand{\tr}{\mbox{tr}}
\newcommand{\la}{\lambda}
\newcommand{\bt}{\beta}
\newcommand{\del}{\delta}
\newcommand{\ep}{\epsilon}
\newcommand{\ta}{\theta}
\newcommand{\ka}{\kappa}
\newcommand{\f}{\phi}
\newcommand{\vf}{\varphi}
\newcommand{\F}{\Phi}
\newcommand{\al}{\alpha}
\newcommand{\ga}{\gamma}
\newcommand{\de}{\delta}
\newcommand{\si}{\sigma}
\newcommand{\Si}{\Sigma}
\newcommand{\za}{\zeta}
\newcommand{\bnabla}{\mbox{\boldmath $\nabla$}}
\newcommand{\bomega}{\mbox{\boldmath $\omega$}}
\newcommand{\bOmega}{\mbox{\boldmath $\Omega$}}
\newcommand{\bsi}{\mbox{\boldmath $\sigma$}}
\newcommand{\bchi}{\mbox{\boldmath $\chi$}}
\newcommand{\bal}{\mbox{\boldmath $\alpha$}}
\newcommand{\bpsi}{\mbox{\boldmath $\psi$}}
\newcommand{\brho}{\mbox{\boldmath $\varrho$}}
\newcommand{\beps}{\mbox{\boldmath $\varepsilon$}}
\newcommand{\bxi}{\mbox{\boldmath $\xi$}}
\newcommand{\bbeta}{\mbox{\boldmath $\beta$}}
\newcommand{\ee}{\end{equation}}
\newcommand{\eea}{\end{eqnarray}}
\newcommand{\be}{\begin{equation}}
\newcommand{\bea}{\begin{eqnarray}}

\newcommand{\ii}{\mbox{i}}
\newcommand{\e}{\mbox{e}}
\newcommand{\pa}{\partial}
\newcommand{\Om}{\Omega}
\newcommand{\om}{\omega}
\newcommand{\vep}{\varepsilon}
\newcommand{\bfph}{{\bf \phi}}
\newcommand{\lm}{\lambda}
\def\theequation{\arabic{equation}}
\renewcommand{\thefootnote}{\fnsymbol{footnote}}
\newcommand{\re}[1]{(\ref{#1})}
\newcommand{\R}{{\rm I \hspace{-0.52ex} R}}
\newcommand{\N}{{\sf N\hspace*{-1.0ex}\rule{0.15ex}%
{1.3ex}\hspace*{1.0ex}}}
\newcommand{\Q}{{\sf Q\hspace*{-1.1ex}\rule{0.15ex}%
{1.5ex}\hspace*{1.1ex}}}
\newcommand{\C}{{\sf C\hspace*{-0.9ex}\rule{0.15ex}%
{1.3ex}\hspace*{0.9ex}}}
\newcommand{\eins}{1\hspace{-0.56ex}{\rm I}}
\renewcommand{\thefootnote}{\arabic{footnote}}

\maketitle


\bigskip

\begin{abstract}
A systematic prescription for constructing Chern-Simons gravities in all odd dimensions is given, and 
it is shown that Chern-Simons gravities in $2n+1$ dimensions admit solutions described by the same lapse function which describes the BTZ black hole in the $n=1$ case. This has been carried out explicitly for $n=1,2,3$.
Moreover, it is seen that these solutions are unique.
\end{abstract}
\medskip
\medskip
\newpage
\section{Introduction}
Chern-Simons gravities (CSG) derived from non-Abelian Chern-Simons (CS) densities in $2+1$ dimensions were proposed by Witten in
Ref.~\cite{Witten:1988hc} and they were extended to all odd dimensions by Chamseddine in Refs.~\cite{Chamseddine:1989nu,Chamseddine:1990gk}. The
CSG systems consist of superpositions of gravitational Lagrangians of all possible higher order gravities
in the given dimensions, each appearing with a precise real numerical coefficient resulting from the calculus.

The construction of Chern-Simons gravities employs the usual non-Abelian (nA) Chern-Simons (CS) densities, the gauge group being chosen suitably. non-Abelian CS densities are defined only in odd dimensions since they are derived $via$ a one-step
descent~\cite{Tchrakian:2017fdw} from the corresponding Chern-Pontryagin (CP) density in one dimension higher, in even dimensions~\footnote{The space in which the CS density is defined can be a Minkowskian spacetime as in the present case, or, it can be a
Euclidean space when the volume integral of the CS density plays the role of the Hopfion charge in a Skyrme-Fadde'ev model.}.

Now the CP density for a nA field is the trace of an antisymmetrised product of $n+1$ curvature tensors, so it follows that the corresponding CS density resulting from the one-step descent is also the trace of the product of the nA curvatures, but now also of the
connections. Thus, unlike the CP density which is gauge invariant, the CS density is gauge variant. This is the salient contrast between CP and CS densities.

The nA CS densities in question, in $d=2n+1$ dimensions, are derived from CP densities in one dimension higher for nA gauge group $SO(2n+2)=SO(d+1)$. Conceretely, we adopt the Dirac gamma matrix reprentations for the
generators
\[
\ga_{ab}=-\frac14\,\ga_{[a}\ga_{b]}\ ,\quad a=1,2,\dots,d+1
\]
for the nA $SO(2n+2)=SO(d+1)$ connections. Since this orthogonal group is even, $2n+2$ dimensional, in addition to $\ga_a$, there exists also a chiral matrix $\ga_{2n+3}=\ga_{d+2}$.
Thus in the construction of the CP and CS densities
we have the option of inserting the chiral matrix $\ga_{2n+3}$ under the trace defining these densities, and anticipating the passage from nA gauge systems to gravitational systems we exercise this
option and insert the chiral matrix under the trace defining these
densities. The reason for this is that the resulting gravitational systems, described by all Einstein-Hilbert (EH) Lagrangians, must be Euler type and not Pontryagin type densities.

Since most of the concrete work here will be carried out for $d=3,5,7$, $i.e.$, for $n=1,2,3$, we list these $SO(d+1)$ CS densities $\Omega_{\rm CS}^{(n)}$
\bea
\Omega_{\rm CS}^{(2)}&=&\vep^{\la\mu\nu}\mbox{Tr}\,\ga_5
A_{\la}\left[F_{\mu\nu}-\frac23A_{\mu}A_{\nu}\right]\label{nACS1}\\
\Omega_{\rm CS}^{(3)}&=&\vep^{\la\mu\nu\rho\si}\mbox{Tr}\,\ga_7
A_{\la}\left[F_{\mu\nu}F_{\rho\si}-F_{\mu\nu}A_{\rho}A_{\si}+
\frac25A_{\mu}A_{\nu}A_{\rho}A_{\si}\right]\label{nACS2}
\\
\Omega_{\rm CS}^{(4)}&=&\vep^{\la\mu\nu\rho\si\tau\ka}
\mbox{Tr}\,\ga_9A_{\la}\bigg[F_{\mu\nu}F_{\rho\si}F_{\tau\ka}
-\frac45F_{\mu\nu}F_{\rho\si}A_{\tau}A_{\ka}-\frac25
F_{\mu\nu}A_{\rho}A_{\si}F_{\tau\ka}\nonumber\\
&&\qquad\qquad\qquad\qquad\qquad\qquad
+\frac45F_{\mu\nu}A_{\rho}A_{\si}A_{\tau}A_{\ka}-\frac{8}{35}
A_{\mu}A_{\nu}A_{\rho}A_{\si}A_{\tau}A_{\ka}\bigg]\,,\label{nACS3}
\eea
which we ill employ in the construction of the Chern-Simons gravitational system in these dimensions.

The main aim of the present work is to construct the balck hole solutions of the Chern-Simons gravities (CSG) concretely in $d=3,5,7$ dimensions, and to show that these are described uniquely by the same lapse function $N(r)$ desribing the Banados
$et.\,al.$~\cite{Banados:1992wn} black hole (BH) in $d=3$. It is natural to conjecture thereafter that the same $N(r)$ desribes the BH solutions to CSG systems in all dimensions $d=2n+1$. 

The technical result reported here is part of a larger scheme where gravitational models arising from Higgs--Chern-Simons (HCS) densities, introduced in~\cite{Tchrakian:2010ar,Radu:2011zy,Tchrakian:2015pka}~\footnote{Like the Chern-Simons densities that result from the 
Chern-Pontryagin (CP) densities in one dimension higher, the HCS densities result from the Higgs--Chern-Pontryagin (HCP) densities in one dimensions higher. The latter, HCP densities, being dimensional descendants of CP densities, they are defined in both odd and even
dimensions.}, are considered~\cite{Tchrakian:2017fdw}. Employing HCS densities, which are defined in all dimensions in contrast to the usual CS densities that are defined in odd dimensions only, enables the construction of ``topological gravities'' in all dimensions.
Given that such gravitional systems result from HCS densities, we have referred to them as HCS gravities (HCSG). (These new gravitational models feature a frame-vector and a scalar field in addition to the metric.)

To date, the HCSG model that is quantitatively studied is the simplest such model in $2+1$ dimensions, in~\cite{Radu:2018fda}. There, it was found that the BTZ metric in $2+1$ dimensions plays a pivotal role, which turns out to be the case in all $2n+1$
dimensions~\footnote{To be reported elsewhere shortly.}, namely that the CS gravity (black hole) solutions play a special role in these theories featuring a frame-vector and a scalar field in addition to the $Vielbein$ (the metric).

Building on the insight gained into such theories, it is planned to study quantitatively black hole solutions to HCSG theories of ``topological gravity'' in even spacetime dimensions, as mapped out in~\cite{Tchrakian:2017fdw}.
Most importantly from the physical viewpoint, it is planned to study the case in $3+1$ dimensions proposed in~\cite{Tchrakian:2017fdw}.

\section{Chern-Simons gravities in $d=3,5,7$ dimensions}
The Chern-Simons gravities (CSG) are described by the most general superposition of higher order gravities, in a given dimension.
We refer to the higher order gravities as $p$-Einsten-Hilbert ($p$-EH) gravities, the integer $p$ specifying the number of times the Riemann curvature appears in the given Lagrangian
${\cal L}^{(p,d)}_{\rm EH}$. Employing the $Vielbeine$ $e_\mu^a$, with $\mu=1,2,\dots,d$ labelling the coordinate indices and $a=1,2,\dots,d$ labelling the frame indices, the $p$-EH Lagrangian is
defined as
\bea
\label{EHpd}
{\cal L}^{(p,d)}_{\rm EH}&=&\vep^{\mu_1\mu_2\dots\mu_{2p}\nu_1\nu_2\dots\nu_{d-2p}}
\vep_{a_1a_2\dots a_{2p}b_1b_2\dots b_{d-2p}}\,e_{\nu_1}^{b_1}e_{\nu_2}^{b_2}\dots e_{\nu_{d-2p}}^{b_{d-2p}}
\ R_{\mu_1\mu_2}^{a_1a_2}R_{\mu_3\mu_4}^{a_3a_4}\dots R_{\mu_{2p-1}\mu_{2p}}^{a_{2p-1}a_{2p}}\,.
\eea
Clearly, in the case $d-2p=0$, the density \re{EHpd} is a total divergence, and non-trivial Lagrangians are only those for $d>2p$. For $p=0$ it is the cosmological term.

Like all purely gravitational systems, the Chern-Simons gravities (CSG) support only black hole (BH) solutions, the best known BH solutions being those found by Banados
$et.\,al.$~\cite{Banados:1992wn} in $d=2+1$ dimensions. It is our aim here to extend that result to the next two odd dimensions, $d=5$ and $d=7$, restricted to the radially symmetric case.

We start from the non Abelian (nA) Chern-Simons (CS) densities \re{nACS1}-\re{nACS3} in $d=3,5,7$ dimensons. To pass from a nA gauge system to gravity we employ
the prescription
\bea
A_{\mu}&=&-\frac12\,\om_{\mu}^{ab}\,\ga_{ab}+\ka\,e_{\mu}^{a}\,\ga_{a,d+1}\quad\Rightarrow\quad
F_{\mu\nu}=-\frac12\left(R_{\mu\nu}^{ab}-\ka^2\,e_{[\mu}^{a}\,e_{\nu]}^{b}\right)\ga_{ab}\,,\label{pass}
\eea
restricting our attention to torsionless models. The prescription \re{pass}
relates the nA connection and curvature $(A_\mu,F_{\mu\nu})$ to the spin-connection and Riemann curvature $(\om_\mu^{ab},R_{\mu\nu}^{ab})$. The constant $\ka$ has dimensions of $L^{-1}$ and is chosen to be real such that the resulting gravitational system be a dS model.
To obtain a AdS model, one just makes the replacement $\ka^2\to-\ka^2$.

Substituting \re{pass} in \re{nACS1}, \re{nACS2} and \re{nACS3} and evaluating the traces yields the corresponing Chern-Simons--Gravitatioonal (CSG) Lagrangians ${\cal L}_{\rm CSG}^{(d)}$ in $d=3,5,7$
\bea
{\cal L}_{\rm CSG}^{(3)}&=&-\ka\,\vep^{\mu\nu\la}\vep_{abc}\left(R_{\mu\nu}^{ab}-\frac23\,\ka^2e_{\mu}^ae_{\nu}^b\right)e_{\la}^c\label{3csg0}\\
&=&-\ka\left({\cal L}_{\rm EH}^{(1,3)}-\frac23\,\ka^2{\cal L}_{\rm EH}^{(0,3)}\right)\label{3csg}\\
{\cal L}_{\rm CSG}^{(5)}&=&\ka\,\vep^{\mu\nu\rho\si\la}\vep_{abcde}\left(\frac34\,R_{\mu\nu}^{ab}\,R_{\rho\si}^{cd}-\ka^2\,R_{\mu\nu}^{ab}\,e_{\rho}^ce_{\si}^d
+\frac35\,\ka^4e_{\mu}^ae_{\nu}^be_{\rho}^ce_{\si}^d\right)e_{\la}^e\label{5csg0}\\
&=&\ka\left(\frac34\,{\cal L}_{\rm EH}^{(2,5)}-\ka^2{\cal L}_{\rm EH}^{(1,5)}+\frac35\,\ka^4{\cal L}_{\rm EH}^{(0,5)}\right)\label{5csg}\\
{\cal L}_{\rm CSG}^{(7)}&=&-\ka\,\vep^{\mu\nu\rho\si\tau\ka\la}\vep_{abcdefg}\bigg(\frac18\,R_{\mu\nu}^{ab}\,R_{\rho\si}^{cd}\,R_{\tau\ka}^{ef}
-\frac14\,\ka^2\,R_{\mu\nu}^{ab}\,R_{\rho\si}^{cd}\,e_{\tau}^ee_{\ka}^f\nonumber\\
&&\qquad\qquad\qquad\qquad+\frac{3}{10}\,\ka^4\,R_{\mu\nu}^{ab}\,e_{\rho}^ce_{\si}^de_{\tau}^ee_{\ka}^f
-\frac17\,\ka^6e_{\mu}^ae_{\nu}^be_{\rho}^ce_{\si}^de_{\tau}^ee_{\ka}^f\bigg)e_{\la}^g\label{7csg0}\\
&=&-\ka\left(\frac18\,{\cal L}_{\rm EH}^{(3,7)}-\frac14\ka^2\,{\cal L}_{\rm EH}^{(2,7)}+\frac{3}{10}\ka^4\,{\cal L}_{\rm EH}^{(1,7)}-\frac17\ka^6\,{\cal L}_{\rm EH}^{(0,7)}\right)\label{7csg}
\eea

The $p$-Einstein-Hilbert Lagrangians ${\cal L}_{\rm EH}^{(p,d)}$ in \re{3csg}, \re{5csg} and \re{7csg} in $d=3,5,7$ are those defined by \re{EHpd}, with $p$ ranging from $p=0$ up to the maximum allowed
$p=d$ in each case. They will be subjected to static radial symmetry and the resulting one
dimensional reduced densities will be subjected to variations. Solving the resulting equations of motion will yield the sought after BH solutions.

The equivalent expressions \re{3csg0}, \re{5csg0}\ and \re{7csg0} for ${\cal L}_{\rm CSG}^{(d)}$ are also useful, in that they afford a transparent expression for the Euler-Lagrange equations
resulting from the variation w.r.t. the $Vielbein$. Employing the short hand notation
\be
\label{shhd}
\bar{R}_{\mu\nu}^{ab}=R_{\mu\nu}^{ab}-\ka^2e^a_{[\mu}e^b_{\nu]}\,,
\ee
the Einstein equations $E^{(n)}{}_a^\mu=0$ resulting from subjecting CSG Lagrangians \re{3csg0}, \re{5csg0}\ and \re{7csg0} to variataions w.r.t. the $Vielbein$ $e_\mu^a$, are
\bea
-\ka\,\vep^{\mu\nu\la}\vep_{abc}\,\bar{R}_{\mu\nu}^{ab}&=&0\label{E3}\\
\frac34\ka\,\vep^{\mu\nu\rho\si\la}\vep_{abcde}\,\bar{R}_{\mu\nu}^{ab}\bar{R}_{\rho\si}^{cd}&=&0\label{E5}\\
-\frac18\ka\,\vep^{\mu\nu\rho\si\tau\ka\la}\vep_{abcdefg}\,\bar{R}_{\mu\nu}^{ab}\bar{R}_{\rho\si}^{cd}\bar{R}_{\tau\ka}^{ef}&=&0\,.\label{E7}
\eea
for $n=1,2,3$. It is clear from \re{E3}, \re{E5} and \re{E7} that the Einstein equation for arbitrary $n$ can be expressed by the remarkably simple expression
\be
E^{(n)}[\bar{R}]_a^\mu\stackrel{\rm def.}=\vep^{\mu\nu_1\nu_2\dots\nu_{2n-1}\nu_{2n}}\vep_{ab_1b_2\dots b_{2n-1}b_{2n}}\bar{R}_{\nu_1\nu_2}^{b_1b_2}\dots\bar{R}_{\nu_{2n-1}\nu_{2n}}^{b_{2n-1}b_{2n}}=0\label{wedge}
\ee 
summarising the extrapolation of \re{E3}-\re{E7}.

An interesting byproduct of the Euler-Lagrange equations \re{wedge} is, that a prescription for constructing the CSG Lagrangian ${\cal L}_{\rm CSG}^{(d)}$
for arbitrary $d=2n+1$ can be given.

To describe this procedure, it is convenient to rescale each ${\cal L}_{\rm CSG}^{(d)}$ such that the highest order term in the Riemann curvature, $R^n$ enters with unit coefficient. Thus
\bea
\bar{\cal L}_{\rm CSG}^{(d)}&=&(-)^n\ka\sum_{p=n}^{p=0}\al_{(p,d)}{\cal L}_{\rm EH}^{(p,d)}\ ,\quad{\rm with}\quad \al_{(n,d)}=1\label{LCSGd1}
\eea
having assigned an arbitrary real coefficient $\al_{(i,d)}\ ,\ \ i=n-1,\,n-2,\dots,0$ to each term featuring the $i$-th power of the Riemann curvature.

Calculating the Euler-Lagrange equations w.r.t. the $Vielbein$ $e_\mu^a$ and identifying the result, $\del_{e_\mu^a}\bar{\cal L}_{\rm CSG}^{(d)}=0$, with the Einstein equation $E^{(n)}{}_a^\mu=0$. \re{wedge}, all the unknown coefficients $\al_{(i,d)}$ can be evaluated.
(This involves the replacement of $R_{\mu\nu}^{ab}$ in $\del_{e_\mu^a}\bar{\cal L}_{\rm CSG}^{(d)}$ by $\bar R_{\mu\nu}^{ab}$, using \re{shhd} and setting the coefficients of all terms except the unit coefficient of the $\bar R^n$ term equal to zero.)

As an example, we quote this result for $\bar{\cal L}_{\rm CSG}^{(9)}$
\be
\al_{(3,9)}=-\frac43\ ,\quad\al_{(2,9)}=\frac65\ ,\quad\al_{(0,9)}=-\frac79\,.\label{ald9}
\ee
The results here elaborate on those given in Appendix {\bf A} of Ref.~\cite{Crisostomo:2000bb}.

\subsection{Spherically symmetric black hole solutions}
We now impose static radial symmetry $via$ the generic spherically symmetric line-element in terms of two unknown functions $N(r)$ and $\si(r)$, with
\begin{eqnarray}
\label{metric}
 ds^2=\frac{dr^2}{N(r)}+r^2 d\Omega^2_{ d-2}-\sigma(r)^2 N(r) dt^2,
\end{eqnarray} 
where $r,t$ are the radial and time coordinates, while
$d\Omega^2_{1,d-2}$ denotes the unit metric on $S^{d-2}$. 

The reduced one dimensional Lagrangians $L_{\rm EH}^{(p,d)}$ pertaining to ${\cal L}_{\rm EH}^{(p,d)}$ defined by \re{EHpd}, which  are systematically calculated in a slightly different normalisation
in Refs.~\cite{Chakrabarti:1999mb,Chakrabarti:2001di,Radu:2009rs}, are given here as
\begin{eqnarray}
\label{Lpd}
L^{(p,d)}_{\rm EH}=(d-2)!(d-2p)\,\sigma(r)\, \frac{d}{dr}[r^{d-2p-1}(1-N(r))^p].
\end{eqnarray}

We now replace each term ${\cal L}_{\rm EH}^{(p,d)}$ in \re{3csg}, \re{5csg} and \re{7csg} by the corresponding reduced one-dimensional $L_{\rm EH}^{(p,d)}$ given by \re{Lpd}.
The result is strikingly simple, yielding the expression for each $(p,d)$,
\be
\label{redLpd}
L_{\rm CSG}^{(d)}\simeq(-1)^{n}\ka\,\si\frac{d}{dr}[(1-N)-2\ka^2r^2]^n\ ,\quad n=1,2,3\,,
\ee
up to a numerical constant depending on $n$, which can be evaluated only when the CSG Lagrangian is given for that $n$. In the cases \re{3csg}, \re{5csg} and \re{7csg} at hand these coefficients are ($3!$),
($\frac92\cdot5!$) and ($15\cdot7!$) for $d=3$, $d=5$ and $d=7$ respectively.

Subjecting \re{redLpd} to variations w.r.t. $\si$, and respectively w.r.t. $N$, result in
\bea
[(1-N)-2\ka^2r^2]^n&=&{\rm const.}\label{sqbr}\\
\left[(1-N)-2\ka^2r^2\right]^{n-1} \,\si'&=&0\,.\label{nsoln}
\eea

Choosing to solve \re{nsoln} by setting $\si'=0$ (or $\si={\rm const.}=1$),
we have the unique solution
\be
\label{soln}
N=c-2\ka^2r^2\ ,\quad\si={\rm const.}=1\,,
\ee
where $c$ is related to the integration constant. The case $c=1$ corresponds to the (A)dS space depending on the sign of $\ka^2$. The solution \re{soln} describes a AdS black hole when $c<0$ and $\kappa^2<0$. (In this case
the square bracket in \re{nsoln} does not vanish and hence $\si'=0$ and $\si=1.$)

In view of the unique result \re{soln} for $d=3,5,7$, it is reasonable to conjecture that the solution \re{soln} holds for all $d=2n+1$. 
These are the black hole solutions of CS gravity in $d=2n+1$ claimed here, which are implicit in Ref.~\cite{Crisostomo:2000bb}. For $d=3$ this is the familiar BTZ solution~\cite{Banados:1992wn}.
                                                                                                                              
In dimensions $d\ge 5$  ($n\ge 2$), \re{nsoln} can be solved by requiring that the square bracket vanishes. In that case $\si(r)$ is not fixed and $c=1$ in \re{soln}. These solutions, which appear only in dimensions $d\ge 5$, are the
'special degenerate vacuum solutions' given in Ref.~\cite{Maeda:2011ii}.


The situation here, where the black holes in all odd dimensions are desribed by the same lapse function $N(r)$ given by \re{soln},
is reminiscent of the unit charge BPST instantons~\cite{Belavin:1975fg} of the Yang-Mills (YM) system on $\R^4$ and the instantons~\cite{Tchrakian:1984gq} of the 
$p$-YM systems on $\R^{4p}$, which are described by the same radial structure function
$w(r)$ for all $p$. This situation holds also for the solitons of the $O(2p+1)$ Skyrme systems on $\R^{2p}$, where the Belavin-Polyakov vortices of the $O(3)$ sigma model on
$\R^2$~\cite{Polyakov:1975yp} and the unit charge Skyrmions of the $O(2p+1)\,\ (p\ge 2)$ Skyrme
 systems on $\R^{2p}$~\cite{Brihaye:2017wqa}, are all described by the same chiral function $f(r)$.

\section{Summary}
The Chern-Simons gravitational (CSG) Lagrangians ${\cal L}_{\rm CSG}^{(d)}$ in dimensions $d=1,2,3$, listed in \re{3csg}, \re{5csg} and \re{7csg}, are expressed in terms of the usual
$p$--Einsten-Hilbert ($p$-EH) Lagrangians ${\cal L}_{\rm EH}^{(p,d)}$ defined by \re{EHpd} in $d$ dimensions. It is shown that the Euler-Lagrange equations, \re{wedge}, can be stated in all $2n+1$ dimensions and this result is exploited to give a prescription for
constructing the CSG Lagrangians in all (odd) dimensions. This complements some of the results in Ref.~\cite{Crisostomo:2000bb}.

Then, each ${\cal L}_{\rm EH}^{(p,d)}$ is subjected to static radial symmetry and
the result is substituted in the corresponding CSG Lagrangian \re{3csg}, \re{5csg} or \re{7csg}. The resulting Euler-Lagrange equations yield the unique black hole solution $N(r)$ \re{soln}, which
is the lapse function describing the BTZ~\cite{Banados:1992wn} black hole. Based on this, it may be reasonable to conjecture that this result holds for all $d=2n+1$.

\bigskip
\noindent
{\bf Acknowledgements}
I am deeply indepted to Eugen Radu for his involvement and help at all stages of this work. I would like to thank Ruben Manvelyan and Grainne O'Brien for very helpful discussions. Special thanks to Amithabha Chakrabarti for his early collaboration on this project.
Thanks are due to Hideki Maeda for bringing Ref.~\cite{Maeda:2011ii} to my attention.

\begin{small}

\end{small}

\begin{thebibliography}{99}
\bibitem{Witten:1988hc}
  E.~Witten,
  ``(2+1)-dimensional gravity as an exactly soluble system,''
  Nucl.\ Phys.\ B {\bf 311} (1988) 46.
\bibitem{Chamseddine:1989nu}
  A.~H.~Chamseddine,
  ``Topological gauge theory of gravity in five-dimensions and all odd dimensions,''
  Phys.\ Lett.\ B {\bf 233} (1989) 291.
\bibitem{Chamseddine:1990gk}
  A.~H.~Chamseddine,
  ``Topological gravity and supergravity in various dimensions,''
  Nucl.\ Phys.\ B {\bf 346} (1990) 213.
\bibitem{Tchrakian:2017fdw}
  D.~H.~Tchrakian,
  Phys.\ Atom.\ Nucl.\  {\bf 81} (2018) no.6,  930
  doi:10.1134/S1063778818060297
  [arXiv:1712.05190 [gr-qc]].
\bibitem{Banados:1992wn}
  M.~Banados, C.~Teitelboim and J.~Zanelli,
  Phys.\ Rev.\ Lett.\  {\bf 69} (1992) 1849
  [hep-th/9204099].
\bibitem{Tchrakian:2010ar}
  T.~Tchrakian,
  J.\ Phys.\ A {\bf 44} (2011) 343001
  doi:10.1088/1751-8113/44/34/343001
  [arXiv:1009.3790 [hep-th]].
\bibitem{Radu:2011zy}
  E.~Radu and T.~Tchrakian,
  doi:10.1142/9789814440349/0020
  arXiv:1101.5068 [hep-th].
\bibitem{Tchrakian:2015pka}
  D.~H.~Tchrakian,
  J.\ Phys.\ A {\bf 48} (2015) no.37,  375401
  doi:10.1088/1751-8113/48/37/375401
  [arXiv:1505.05344 [hep-th]].
\cite{Radu:2018fda}
\bibitem{Radu:2018fda}
  E.~Radu and D.~H.~Tchrakian,
  Class.\ Quant.\ Grav.\  {\bf 35} (2018) no.17,  175012
  doi:10.1088/1361-6382/aad3da
  [arXiv:1804.09902 [gr-qc]].
\bibitem{Crisostomo:2000bb}
  J.~Crisostomo, R.~Troncoso and J.~Zanelli,
  Phys.\ Rev.\ D {\bf 62} (2000) 084013
  doi:10.1103/PhysRevD.62.084013
  [hep-th/0003271].
\bibitem{Chakrabarti:1999mb}
  A.~Chakrabarti and D.~H.~Tchrakian,
  Adv.\ Theor.\ Math.\ Phys.\  {\bf 3} (1999) 791
  doi:10.4310/ATMP.1999.v3.n4.a2
  [hep-th/9908128].
\bibitem{Chakrabarti:2001di}
  A.~Chakrabarti and D.~H.~Tchrakian,
  Phys.\ Rev.\ D {\bf 65} (2002) 024029
  doi:10.1103/PhysRevD.65.024029
  [hep-th/0101160].
\bibitem{Radu:2009rs}
  E.~Radu and D.~H.~Tchrakian,
  arXiv:0907.1452 [gr-qc].
\bibitem{Maeda:2011ii}
  H.~Maeda, S.~Willison and S.~Ray,
  Class.\ Quant.\ Grav.\  {\bf 28} (2011) 165005
  doi:10.1088/0264-9381/28/16/165005
  [arXiv:1103.4184 [gr-qc]].
\bibitem{Belavin:1975fg}
  A.~A.~Belavin, A.~M.~Polyakov, A.~S.~Schwartz and Y.~S.~Tyupkin,
  Phys.\ Lett.\ B {\bf 59} (1975) 85
   [Phys.\ Lett.\  {\bf 59B} (1975) 85].
  doi:10.1016/0370-2693(75)90163-X
\bibitem{Tchrakian:1984gq}
  D.~H.~Tchrakian,
  Phys.\ Lett.\  {\bf 150B} (1985) 360.
  doi:10.1016/0370-2693(85)90994-3
\bibitem{Polyakov:1975yp}
  A.~M.~Polyakov and A.~A.~Belavin,
  JETP Lett.\  {\bf 22} (1975) 245
   [Pisma Zh.\ Eksp.\ Teor.\ Fiz.\  {\bf 22} (1975) 503].
\bibitem{Brihaye:2017wqa}
  Y.~Brihaye, C.~Herdeiro, E.~Radu and D.~H.~Tchrakian,
  JHEP {\bf 1711} (2017) 037
  doi:10.1007/JHEP11(2017)037
  [arXiv:1710.03833 [gr-qc]].








\end{thebibliography}
\end{document}